\documentclass[
 aps,
 11pt,
 final,
 notitlepage,
 oneside,
 twocolumn,
 nobibnotes,
 nofootinbib,
 superscriptaddress,
 centertags]
{revtex4}

\def\saoname{Special Astrophysical Observatory,  Russian Academy of Sciences,
              Nizhnii Arkhyz, 369167 Russia}

\def\squareforqed{\hbox{\rlap{$\sqcap$}$\sqcup$}}

\def\sq{\ifmmode\squareforqed\else{\unskip\nobreak\hfil
\penalty50\hskip1em\null\nobreak\hfil\squareforqed
\parfillskip=0pt\finalhyphendemerits=0\endgraf}\fi}

\def\sun{\hbox{$\odot$}}

\def\arcmin{\hbox{$^\prime$}}

\def\arcsec{\hbox{$^{\prime\prime}$}}

\def\utw{\smash{\rlap{\lower5pt\hbox{$\sim$}}}}

\def\udtw{\smash{\rlap{\lower6pt\hbox{$\approx$}}}}

\def\diameter{{\ifmmode\mathchoice
{\ooalign{\hfil\hbox{$\displaystyle/$}\hfil\crcr
{\hbox{$\displaystyle\mathchar"20D$}}}}
{\ooalign{\hfil\hbox{$\textstyle/$}\hfil\crcr
{\hbox{$\textstyle\mathchar"20D$}}}}
{\ooalign{\hfil\hbox{$\scriptstyle/$}\hfil\crcr
{\hbox{$\scriptstyle\mathchar"20D$}}}}
{\ooalign{\hfil\hbox{$\scriptscriptstyle/$}\hfil\crcr
{\hbox{$\scriptscriptstyle\mathchar"20D$}}}}
\else{\ooalign{\hfil/\hfil\crcr\mathhexbox20D}}%
\fi}}

\begin{document}
\selectlanguage{english}



%


\title{The Stellar Population and Orbit of the Galactic Globular Cluster Palomar~3}

\author{\firstname{M.~E.}~\surname{Sharina}}
\email{sme@sao.ru}
\affiliation{\saoname}
\author{\firstname{M.~V.}~\surname{Ryabova}}
\affiliation{Southern Federal University, Rostov-on-Don, Sadovaya str.105/42, 344006, Russia}

\author{\firstname{M.~I.}~\surname{Maricheva}}
\affiliation{Kazan Federal University, Kremlevskaya str. 18, Kazan, 420008 Russia}
 
 \author{\firstname{A.~S.}~\surname{Gorban}}
\affiliation{Southern Federal University, Rostov-on-Don, Sadovaya str.105/42, 344006, Russia}
 
\begin{abstract}
Deep stellar photometry of one of the most distant Galactic globular clusters, Palomar~3, based
on frames taken with the VLT in Johnson-Cousins broadband V and Ifilters is presented, together with
medium-resolution stellar spectroscopy in the central region of the cluster obtained with the CARELEC
spectrograph of the Observatoire de Haute Provence and measurements of the Lick spectral indices for
the integrated spectrum. Computations of the orbital parameters of Palomar~3 and nine Galactic globular
clusters with similar metallicities and ages are also presented. The orbital parameters, age, metallicity,
and distance of Palomar~3 are estimated. The interstellar absorption is consistent with and supplements
values from the literature. The need to obtain more accurate data on the proper motions, ages, and chemical
compositions of the cluster stars to elucidate the origin of this globular cluster is emphasized.
\end{abstract}


\maketitle

\section{INTRODUCTION}
\label{intro}

\begin{table*}[]
\setcaptionmargin{0mm} \onelinecaptionstrue \captionstyle{normal}
\caption{Main properties of Pal~3 from the literature: (1, 2) right ascension and declination, (3) color excess in
magnitudes, (4) metallicity [dex], (5) distance from the Sun in kpc, (6, 7) apparant and absolute magnitudes, (8) central
surface brightness in mag/arcsec2 , (9) half-luminosity radius in pc, (10) heliocentric radial velocity in km/s, (11) $\rm HBR$,
and (12) age in billions of years. Thefirst row gives values from \cite{Hilker} and the second row data from the catalog \cite{Harris}.}
\label{propGCs}
\medskip
\begin{tabular}{c|c|c|c|c|c|c|c|c|c|c|c}
\hline
RA(2000)   & DEC(2000) & $\rm E(B-V)$ & $\rm [Fe/H]$& $\rm Dist$&  $\rm V_{t}$ & $\rm \mu_V$& $\rm r_h$ &  $\rm M_V$ & $\rm V_h$ & HBR & T\\
(1)        &  (2)      &  (3)         &  (4)        &  (5)      &  (6)         & (7)        & (8)       & (9)         & (10)     & (11)& (12)\\
\hline                                                                            
10:05:31.9 & +00:04:18  &  0.04    & -1.74 &   91.9     &  14.91   & 23.84 & 19.3  & -5.03 & --   & --   & 10    \\
10:05:31.4 & +00:04:17  &  0.04    & -1.66 &   92.7     &  14.26   & 23.08 & 17.8  & -5.70 & 83.4 & -0.5 & -- \\
\hline
\end{tabular}
\end{table*}
Like other Palomar globular clusters \cite{Harris}, Palomar~3 (Pal~3)
was discovered in the 1950s on photographic plates of the Palomar Sky Survey \cite{Abell}. The
main observational characteristics of this cluster from the literature are listed in Table~\ref{propGCs}.
Pal~3 is $\sim$96~kpc from the Galactic center (\cite{Harris}, \cite{Marsakov}, \cite{Carretta})
in the Galactic halo, and is not a member of any known stellar streams,
such as the Sagittarius stream (\cite{Palma}, \cite{Bellazzini}).

There are only seven globular clusters at distances of 40--120 kpc from the Galactic center \cite{Harris}: AM 1,
Eridanus, Pyxis, Pal~3, Pal~4, Pal~14, and NGC~2419. Their origin is not clear (see, e.g., \cite{f17}). Apart from
NGC~2419, all have red horizontal branches, low
stellar densities, and similar masses and metallicities (\cite{Harris}, \cite{go97}). 
The range of their masses is $\rm (1.8\div6) \cdot 10^4 M_{\sun}$ \cite{go97}, and the range of their metallicities is
$\rm-1.3<[Fe/H]<-1.8$~dex\footnote{The iron content in solar units is $\rm[Fe/H]=log(N_{Fe}/N_H)-log(N_{Fe}/N_H)_{\sun}$,
 where $\rm N_{Fe}/N_H$ is the ratio of the abundances of iron and hydrogen in terms of numbers of atoms, or in
terms of mass, which is related to the mass fraction of elements heavier than helium $Z$ by an empirical formula (see,
e.g., \cite{Bertelli94}). The solar mass fractions of hydrogen $X$, helium $Y$,
and metals $Z$ are given in \cite{Asplund}. Obviously, $X+Y+Z=1$.}. 
NGC~2419 is a more massive object ($\rm 1.6 \cdot10^6 M_{\sun} $) with a low metallicity
$\rm [Fe/H]<-2.1$~dex and a blue horizontal branch.

The first estimate of the metallicity of Pal~3 was obtained by Gratton and Ortolani \cite{go84} based on the
object's color-magnitude diagram (CMD): $\rm [Fe/H]=-1.4\pm0.4$ (this value is $\rm [Fe/H]=-1.57\pm0.19$ \cite{lee94})
on the scale of Zinn and West \cite{zw84}). Subsequent estimates of the metallicity of Pal~3 based on its
CMD range from $\rm [Fe/H]=-1.4 \div -1.7$  (\cite{Hilker} and references therein). Lee et al. \cite{lee94} used the CMD of
the cluster to determine the horizontal branch ratio, (\mbox{$\rm HBR = (B-R)/(B+V+R)$}), which serves as an
index characterizing the numbers of stars in different parts of the horizontal branch, which they found to
be -0.82 for Pal~3. Harris \cite{Harris} presents another value in his catalog (Table~\ref{propGCs}). Catelan et al. \cite{Catelan} found
that the dispersion of the masses of horizontal-branch stars in Pal~3 was significantly lower than for the
cluster M3. According to estimates of the age of Pal~3 based on its CMD (\cite{Stetson}, \cite{Vandenberg}, \cite{Hilker}, \cite{Dotter08}), 
this cluster is approximately one to two billion years younger than clusters in the Galaxy belonging to the old halo
subsystem, such as M3 and M13\footnote{There is currently no unique system for dividing globular
clusters among various Galactic subsystems (see, e.g., \cite{Carretta}, \cite{Marsakov}). 
It is believed that clusters of the young halo were members of dwarf galaxies in the past. Such clusters are
$\sim$1-2 billion years younger than objects of the old halo (13--14 billion years). Absolute ages in billions of years are
known for only some objects. Only relative ages are known for most globular clusters (e.g., \cite{MarinFranch}). 
Clusters of the disk rotate at a modest height above the Galactic plane ($\leq1$~kpc)
with roughly the same rotational speed as the disk, and are, on average, younger than old-halo objects. Clusters of the
bulge have ages comparable to those of old-halo objects, but are located near the Galactic center, at Galactocentric
distances of \mbox{$ R_{gc}\leq3-4$}~kpc. Globular clusters of the halo are conventionally separated into 
inner-halo and outer-halo objects, with the tentative division between them being at $ R_{gc}=15$~kpc \cite{Carretta}. 
Clusters located in the young and old halo can be members of either the outer or the inner halo. 
Clusters located in the bulge can belong to the inner halo.}

Borissova et al. \cite{Borissova}  discovered a Type~II Cepheid in Pal~3. Such Cepheids are usually encountered 
in massive globular clusters with blue horizontal branches. Based on his deep CCD photometry of stars in Pal~3 with 
the Very Large Telescope (VLT, Paranal, Chile) in Johnson-Cousins broadband B and Vfilters, Hilker \cite{Hilker} 
distinguished stars that were probable cluster members and estimated the cluster's metallicity and age (Table~\ref{propGCs}).
Sohn et al. \cite{Sohn} carried out photometry of stars in the broad vicinity of Pal~3 using large-scale frames of
a mosaic CCD of the Canada-France-Hawaii telescope. They detected an elongation in the distribution
of the cluster stars along the direction of the Galactic anti-center.

There is no unified opinion about whether or not Pal~3 belongs to the young or the old halo. 
Koch et al. \cite{Koch} determined the elemental abundances for four red giants in the cluster via high-resolution
spectroscopy. They also presented the results of the best deep B and V photometry of Pal~3 currently
available, obtained with the LRIS instrument on the Keck~I telescope; the main-sequence turn-off point
is just barely visible in these data. Koch et al. \cite{Koch} fitted the observed distribution of stars in the CMD
using an isochrone with an enhanced abundance of alpha-process elements ($\rm [\alpha/Fe]$), $\rm[Fe/H]=-1.6$~dex
and an age of 10 billion years, developed by Pietrinferni et al. \cite{Pietrinferni}. Based on high-resolution spectra
obtained for four of the cluster stars and model stellar atmospheres, it was concluded that the metallicity of
the cluster indicated by the Ca~II lines of the CaT (8498, 8542 and 8662 \AA) and the Mg~I lines
was $\rm[Fe/H]=-1.58\pm0.02(stat.)\pm0.13(syst.)$~dex.
Koch et al. \cite{Koch} also concluded that the abundances of the alpha-process elements Mg, Si, and Ca and
of iron-peak and neutron-capture elements were enhanced. This same picture was observed for old halo
globular clusters with the same metallicity. Koch et al. \cite{Koch} suggested that the chemical composition
of Pal~3 does not resemble those for stars in dwarf galaxies. The heavy elements in Pal~3 were produced
mainly in slow nucleosynthesis processes. Koch et al. \cite{Koch} did not detect any variations in the elemental
abundances in the cluster stars.

Palma et al. \cite{Palma} proposed that Pal~3 was earlier a member of the Phoenix dwarf irregular galaxy, since
the orbit of the cluster is coplanar with the orbit of this galaxy. Pal~3 is now located near the pericenter
of its orbit, with the apocenter located at a distance of 445 kpc \cite{Palma}. This distance coincides with the distance
to the Phoenix galaxy.

Orbits for Pal~3 were computed by Dinescu et al. \cite{Dinescu} and Balbinot and Gieles \cite{Balbinot}. Both analyses
used the proper motions of \cite{mc93}, but the two obtained different orbital parameters.

In our current study, we have carried out photometric and spectroscopic studies of Pal~3, and also
computed the orbit of this object. This study was motivated by the lack of a unified opinion about the
structure of the horizontal branch, the metallicity, and the age of this cluster, which are required to
study the relationship between the morphology of the horizontal branch and the cluster's evolutionary parameters. 
Another motivation was the contradictory conclusions about the character of the motion and the
origin of Pal~3 encountered in the literature.

\begin{figure*}[]
 \setcaptionmargin{5mm} \onelinecaptionstrue \captionstyle{normal}
 \includegraphics[scale=0.6]{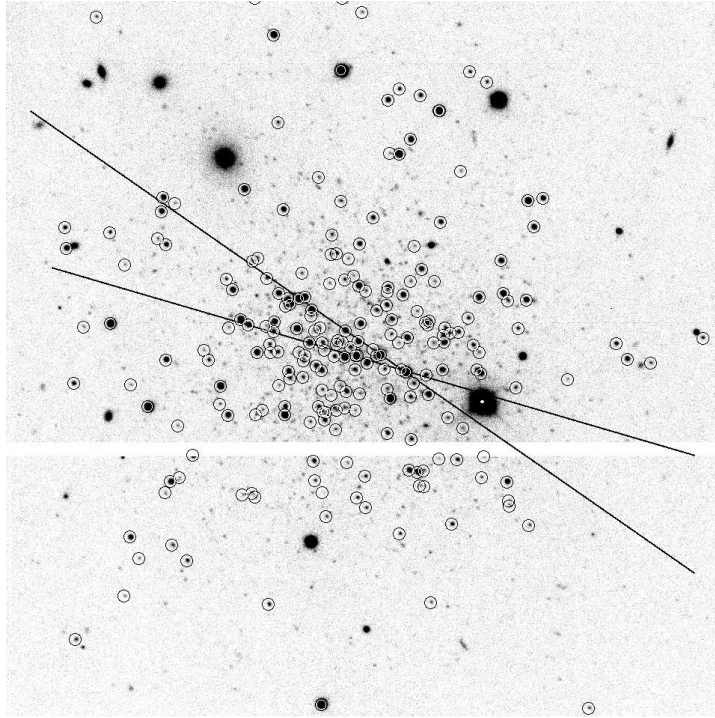}    
 \caption{Illustration of the alignments of the slits. A $4\arcmin \times 4\arcmin$ fragment of a VLT frame in the V filter is shown (North is up, East
is to the right). Stars for which photometry was carried out are indicated.}
\label{slits}
\end{figure*}
\begin{table*}[]
\setcaptionmargin{0mm} \onelinecaptionstrue \captionstyle{normal}
\caption{Journal of VLT observations. The columns give for the two chips the right ascension and declination of the
centers of the CCD frames, exposure time in seconds for the V and I frames, mean FWHM of stellar images in arcseconds
for the V and I frames, and the air masses for the V and I observations.}
\label{logVLT}
\medskip
\begin{tabular}{l|c|c|c c|c c|c c}
\hline
Chip& RA(2000)    & DEC(2000)   & \multicolumn{2}{c}{$T_{exp}$}   & \multicolumn{2}{c}{FWHM}  & \multicolumn{2}{c}{Airmass}  \\ 
\cline{4-5} \cline{6-7} \cline{8-9}
& hh:mm:ss    & gr:mm:ss    & V         & I      & V      & I           &  V      & I \\
\hline
1  & 10:05:31.4  & +00:06:25.9 &  30       &  30    & 1.0    & 0.88        &  1.205  & 1.201\\
   &             &             &  0.3      &  0.3   & 0.75   & 1           &  1.207 & 1.202\\
2  & 10:05:31.4  & +00:02:26.0 &  30       &  30    & 0.98   & 0.95        &  1.205  & 1.201\\
   &             &             &  0.3      &  0.3   & 0.97   & 0.85        &  1.207  & 1.202\\
\hline
\end{tabular}
\end{table*}

\begin{table}[]
\setcaptionmargin{0mm} \onelinecaptionstrue \captionstyle{normal}
\caption{Journal of CARELEC spectral observations (Observatoire de Haute Provence). The columns give the object,
RA and DEC, date, exposure, and the seeing (the mean FWHM of stellar images in arcsec).}
\label{logOHP}
\scriptsize
\medskip
\begin{tabular}{l|c|c|c|c|c}
\hline
Object  & RA(2000)  & DEC(2000)& Date      & $T_{exp}$ & {\tiny FWHM} \\
         & hhmmss    & grmmss  & dd.mm.yy  & (sec.)    & ($\arcsec$) \\
\hline
  Pal~3  & 100531.4  & +000417 & 02.12.08 & 3x1800, & 2.7 \\
         &           &         &          & 900     &      \\
  Pal~3  & 100531.4  & +000417 & 03.12.08 & 4x1800  & 2.5 \\
HD93521  & 104823.5  & +373413 & 02.12.08 & 15      & 2.7  \\
HR1544   & 045036.7  & +085401 & 01.12.08 & 2       & 2.6 \\
  HR1805 & 052419.8  & +342607 & 02.12.08 & 2       & 2.7 \\
         &           &         & 01.12.08 & 2       & 2.6 \\
HR2002   & 054556.7  & +243309 & 01.12.08 & 2       & 2.6 \\
HR2600   & 065538.5  & +380723 & 02.12.08 & 5       & 2.7  \\
HR3418   & 083608.7  & +033105 & 03.12.08 & 5       & 2.5 \\
HR3422   & 083734.2  & +460039 & 01.12.08 & 2       & 2.6 \\
HR3905   & 094955.4  & +261436 & 03.12.08 & 2       & 2.5 \\
HD201626 & 210748.3  & +262438 & 02.12.08 & 20      & 2.7  \\
HR8924   & 232657.0  & -044819 & 01.12.08 & 1       & 2.6 \\
\hline
\end{tabular}
\end{table}

\section{ CHARACTERISTICS OF THE OBSERVATIONAL DATA}
\label{data}

The frames from the archive of the European Southern Observatory
(ESO)\footnote{\url{http://archive.eso.org/cms.html#PRODUCTS}} used for our photometry were obtained on February 23, 2006 on
the Very Large Telescope (Paranal, Chile). Pal~3 was observed with exposures of 0.3 and 30~s in
the V and I filters using the camera of the FORS2 spectrograph. The total size of the two-chip CCD
array was $4000\times4000$  pixels $0.25\times0.25$ in size. A journal of the observations is given in Table~\ref{logVLT}.

The spectral data for Pal~3 were kindly presented by E. Davoust. The spectral observations were
obtained using the CARELEC spectrograph \cite{Lemaitre} mounted on the 1.93-m telescope of the Obser-
vatoire de Haute Provence (OHP). A journal of these observations is given in Table~\ref{logOHP}. The slit
size was 5.5\arcmin x2". A grating with 300 lines/mm was used, which provided a spectral resolution of
approximately 1.78~\AA/pixel over the spectral range 3700--6800~\AA.
Exposures of He and Ne lamps were obtained at the beginning and end of each night,
for subsequent calibration of the wavelength scales of the spectra. Spectrometric standards and stars
from the list of Worthey \cite{Worthey94} (so-called Lick standards) were observed over the course of the night
to enable translation of the measured absorption indices into the standard Lick system and control
of the radial-velocity measurements of the observed globular clusters. Table~\ref{logOHP} lists the standards observed
on December 1-3, 2008. The stars HD93521 and HR1544 are spectrophotometric standards.

\section{REDUCTION OF THE OBSERVATIONAL DATA}
\label{sec_reduc}

The ESO archive contains direct frames that have passed through a standard preliminary reduction, 
including subtraction of the electron zero level, correction for dark current, and flat fielding. 
We removed cosmic-ray traces using the MIDAS program package \cite{b83} {\it filter/cosmic}.

Our stellar photometry was carried out using the DAOPHOT-II program in the MIDAS package \cite{s87}.
The seeing in arcseconds estimated by fitting two-dimensional Gaussians to stellar images is presented
in Table~\ref{logVLT}. We constructed growth curves for roughly a dozen isolated stars with medium brightnesses in
each frame, to determine the aperture corrections. These corrections were roughly 0.4$^m$ lower for the I
than for the V frames, due to the different forms of their point spread functions. The magnitudes obtained 
as a result of applying the aperture corrections were corrected for atmospheric extinction, reduced to
a 1~s exposure time, and translated to the international Johnsons-Cousins photometric system using
the zero points and color factors presented at the ESO site\footnote{\url{ http://www.eso.org/observing/dfo/quality/
FORS2/qc/photcoeff/photcoeffs fors2.html}}.

The alignment of the CARELEC spectrograph slits during the spectral observations is shown in
Fig.~\ref{slits}. The process used to reduce the spectral observations was analogous to the one described, for
example, in \cite{khamid14a}. The reduction of the long-slit spectra was carried out in the MIDAS \cite{b83} and IRAF 
\cite{Tody} packages. The dispersion relation provided a mean wavelength calibration accuracy of about 0.16\AA.
The one-dimensional spectra were extracted using the IRAF procedure {\it apsum}. We removed sky emission
lines using the IRAF procedure {\it background}.

\begin{figure*}[]
 \setcaptionmargin{5mm} \onelinecaptionstrue \captionstyle{normal}
 \begin{tabular}{p{0.48\textwidth}p{0.48\textwidth}}
 \hspace{-1.2cm}
 \includegraphics[scale=0.43,angle=-90]{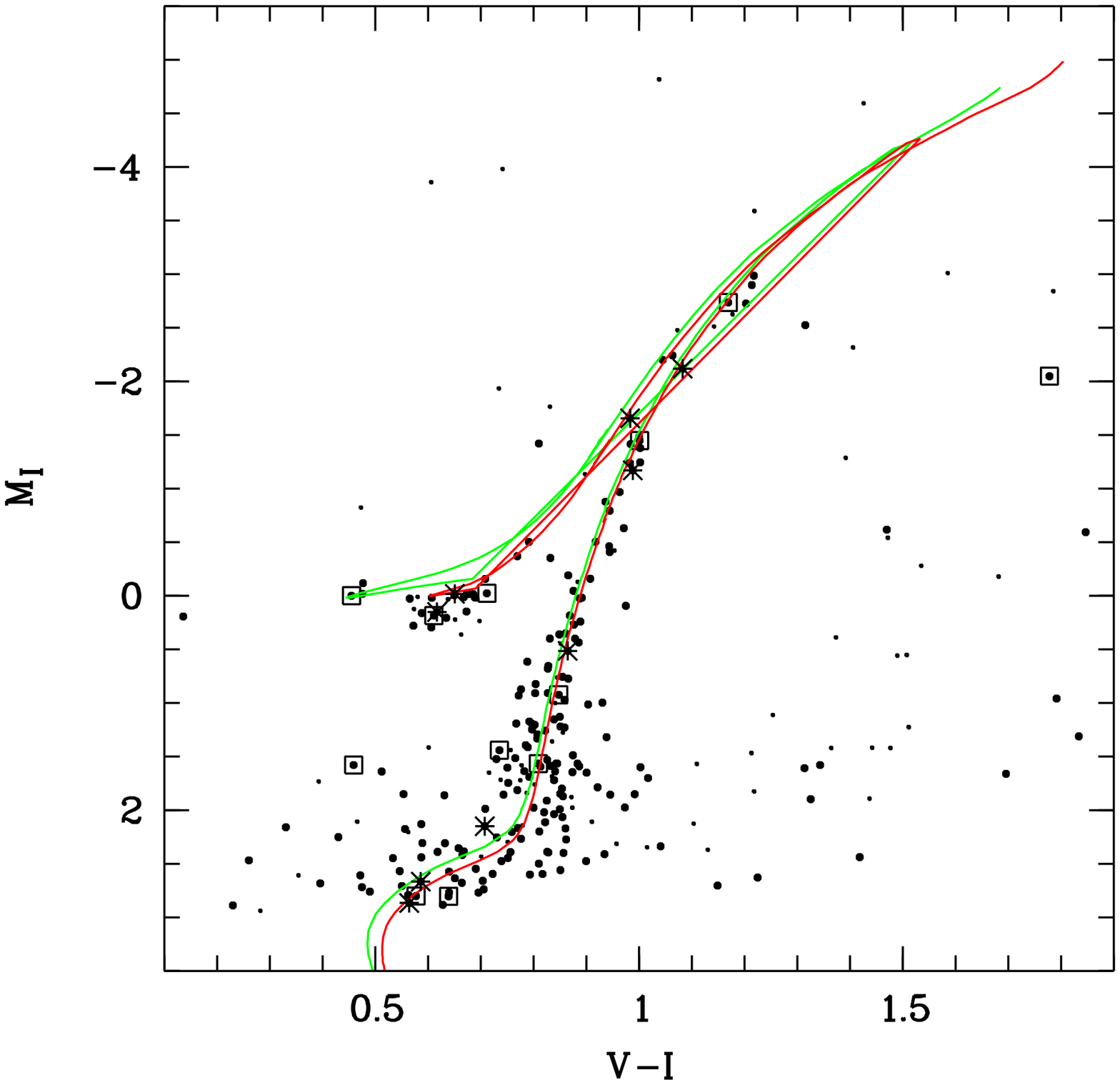}  &
 \hspace{-0.6cm}
 \includegraphics[scale=0.43,angle=-90]{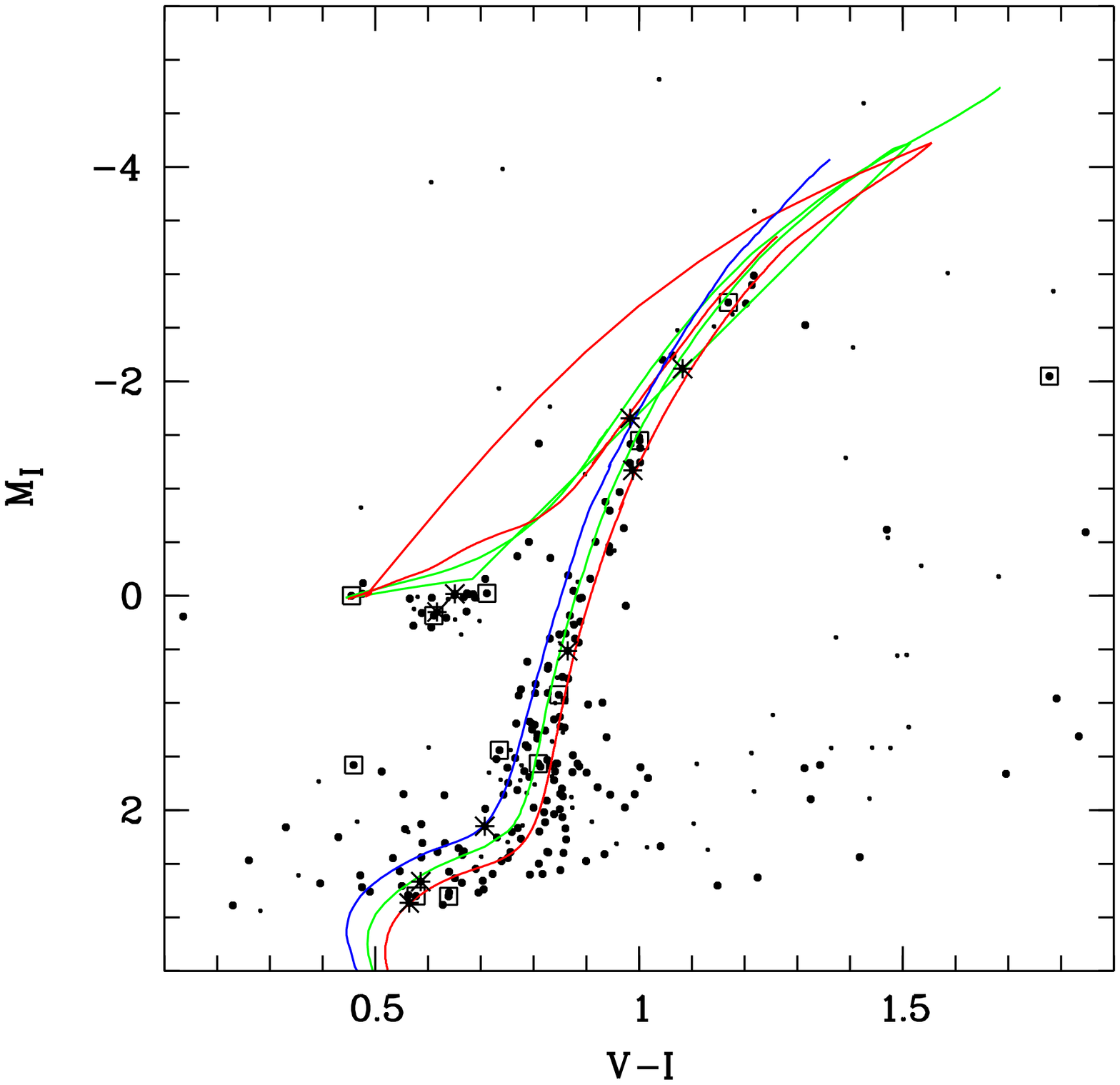}  \\
 \end{tabular}
 \caption{ Color-magnitude diagram for Pal~3 constructed using the VLT photometry data. Various isochrones are shown in
the two panels. The left panel shows the isochrones from \cite{Bertelli} for (1) $\rm Z=0.0004$, $\rm Y=0.23$, $ \rm T=11.2$~billion years (green)
and (2) $\rm Z=0.001$, $\rm Y=0.26$, $\rm T=10$~billion years (red). The left panel shows (1) the isochrone from \cite{Bertelli} for 
$\rm Z=0.0004$, $\rm Y=0.23$, $\rm  T=11.2$~billion years (green), (2) the isochrone of Kim et al. \cite{kim02} for 
$\rm Z=0.0004$, $\rm [\alpha/Fe]=0.3$, $\rm T=10$~billion years
(the red-giant branch passes to the left of isochrone (1)), and (3) the isochrone of Pietrinferni et al. \cite{Pietrinferni} 
$\rm Z=0.0004$, $\rm [\alpha/Fe]=0.3$, $\rm T=10$~billion years (the red-giant branch passes to the right of isochrone (1)). 
Stars within 1.5 of the cluster center are shown by large filled circles, and the remaining stars in the field by small circles. 
The squares and asterisks denote
stars that fell into the two spectrograph slits during the OHP observations (Fig.~\ref{slits}).}
 \label{cmd}
\end{figure*}
\begin{figure*}[]
 \setcaptionmargin{5mm} \onelinecaptionstrue \captionstyle{normal}
 \begin{tabular}{p{0.5\textwidth}p{0.5\textwidth}}
 \includegraphics[scale=0.35,angle=-90]{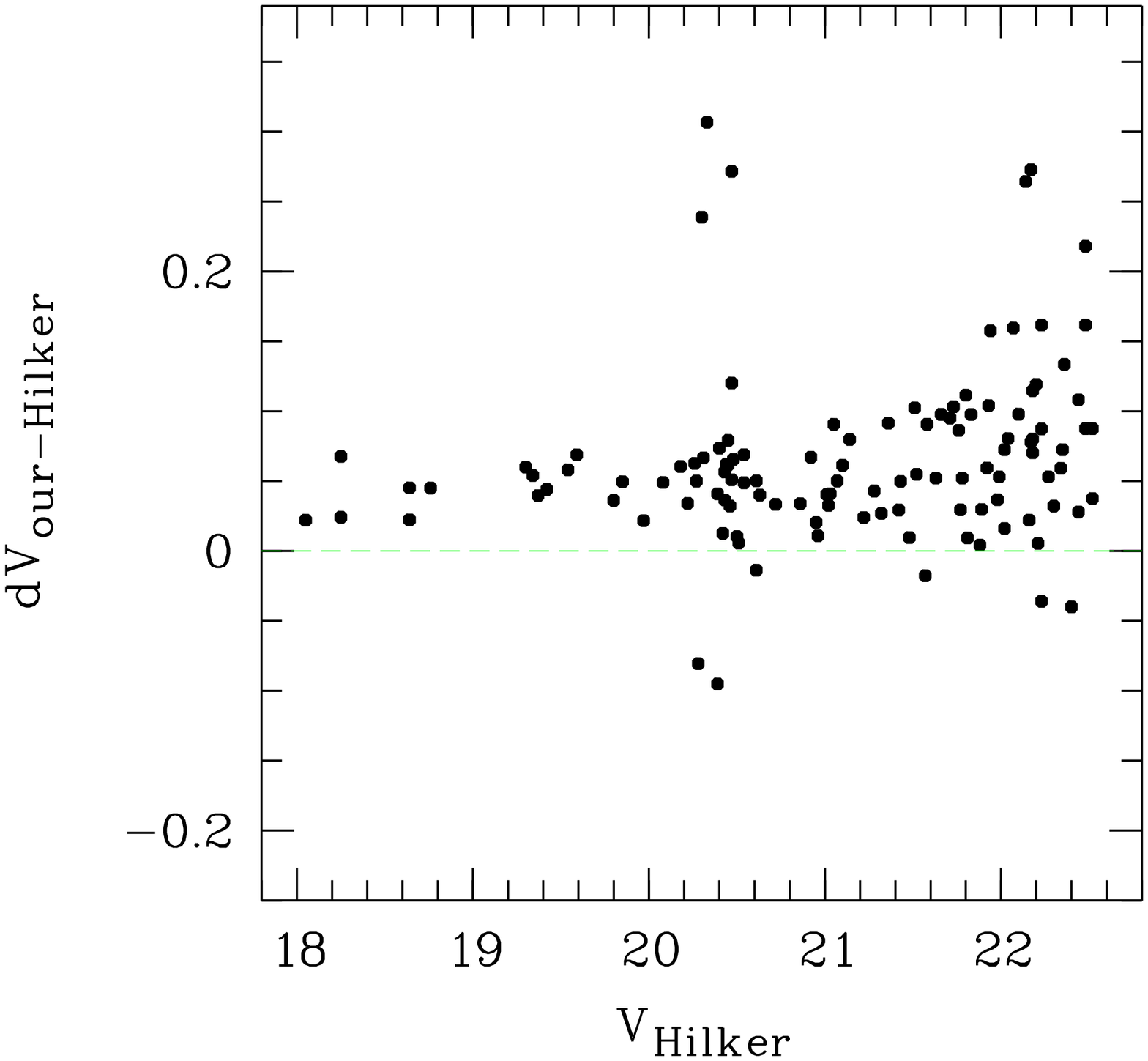} &
 \includegraphics[scale=0.35,angle=-90]{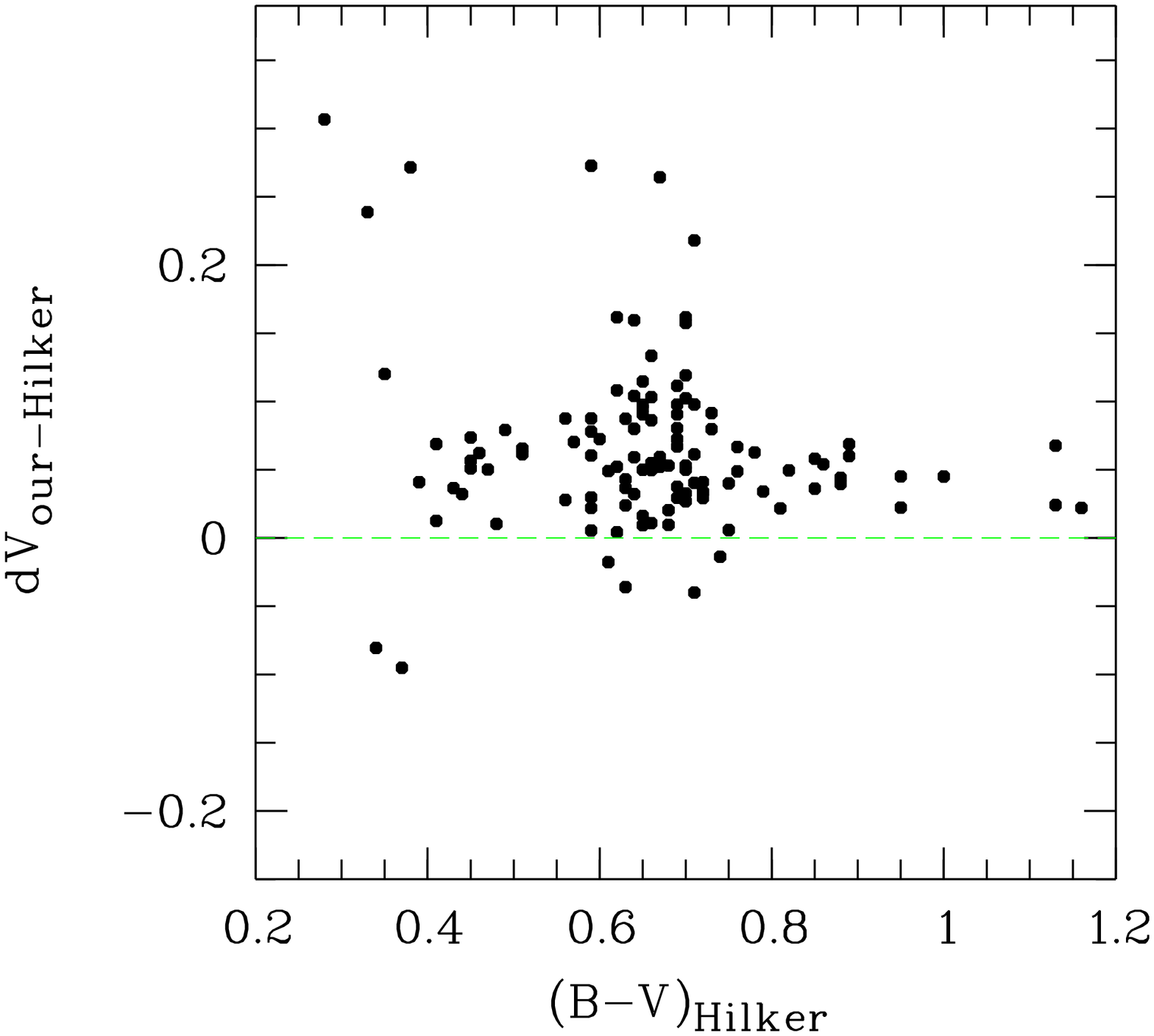} \\
 \end{tabular}
 \caption{Comparison of our photometry of stars in the field of Pal~3 with the results of  \cite{Hilker}.}
 \label{dVHilker}
\end{figure*}

\section{PHOTOMETRIC RESULTS}
\label{photometry}
A table with the results of our photometry of stars in the field of Pal~3 in the VLT frames is presented at
the site ftp: \url{ftp://ftp.sao.ru/pub/sme/Pal3}. 
The CMD of the cluster obtained using these stellar photometry results is shown in Fig.~\ref{cmd}. This diagram has
been constructed using the absolute I magnitude and \mbox{$\rm (V-I)_0$} , corrected for absorption in our 
Galaxy\footnote{$\rm (V-I)_0=(V-I)-E(V-I)$, 
where the color excess is given by $\rm E(V-I)=A_V - A_I$,  and $\rm A_V$ and $\rm A_I$ 
are the absorptions in magnitudes in the V and I filters.}.
The large symbols show stars within the visible boundary of the cluster (within 1.5\arcmin\  of the center of Pal~3).
The stars falling into slits "1"\ and "2" of the spectrograph are shown by different symbols. The left panels shows in 
two different colors the two isochrones from \cite{Bertelli} that best fit the distribution of stars in the
cluster CMD. The optimal position of the isochrone was found using the method described in Section 3
of \cite{khamid14a}. The isochrone of \cite{Bertelli} corresponding to an
age of 11.2 billion years and metallicity $Z=0.0004$  ($\rm [Fe/H]\sim -1.74$,
 according to the empirical calibration of \cite{Bertelli94}), has parameters similar to those of the
isochrone used by Hilker \cite{Hilker} to describe his CMD for the B and V filters.

The right-hand panel of Fig.~\ref{cmd} presents three isochrones. The isochrone of \cite{Bertelli94} with 
$\rm Z=0.0004$, $\rm Y=0.23$, and $ T=11.2$~billion years is repeated in light gray (green in the electronic version). The
isochrone computed by Kim et al.\cite{kim02} with $\rm Z=0.0004$, $\rm [\alpha/Fe]=0.3$, $\rm T=10$~billion years used
by Hilker \cite{Hilker} is shown by the black curve (dark blue in the electronic version). Finally, the isochrone
computed by Pietrinferni et al. \cite{Pietrinferni} with $\rm [Fe/H]=-1.6$~dex, $\rm [\alpha/Fe]=0.3$, 
and $\rm T=10$~billion years used
by Koch et al. \cite{Koch} is shown in dark gray (red in the electronic version). We can see that the red-giant and
subgiant branches and the main-sequence turn-off points for the three isochrones are shifted relative to
one another along the color (temperature) axis. There is no horizontal branch in the isochrones of Kim et al.
\cite{kim02}. The positions of the horizontal branches are close for the isochrones with 
$\rm Z=0.0004$, $\rm Y=0.23$, and $ T=11.2$~billion years [33] and with $\rm [Fe/H]=-1.6$~dex, $\rm [\alpha/Fe]=0.3$, 
and $\rm T=10$~ billion years \cite{Pietrinferni}.

Hilker~\cite{Hilker} used other ESO data and the isochrones computed by Kim et al.~\cite{kim02}. They observed three
exposures in each of their filters ($\rm T_{exp}=25$~s in B and 10 s in V), enabling effective removal of cosmic-ray
traces and fairly deep photometry. The seeing was slightly better during the observations of Hilker~\cite{Hilker}
than in our observations; his stellar photometry was roughly 1m deeper in V than our results (Fig.~\ref{cmd} in this
paper and Fig.~4 in \cite{Hilker}). Nevertheless neither study was able to confidently resolve stars at the 
main-sequence turn-off point in Pal~3. Therefore, the results for the integrated spectrum of the cluster presented
in the previous section, which yield independent estimates of the age and metallicity of the cluster, are
needed to better estimate the distance to Pal~3 and the interstellar absorption in the direction of the cluster.
In both our CMD and the CMD of Hilker~\cite{Hilker}, there is a deficit of bright stars near the tip of the red-giant
branch. Figure 3 shows the difference between our V photometry and the data of Hilker~\cite{Hilker} as a function
of magnitude (left) and color (right). The dispersion of the differences $\rm dV_{our-Hilker}$ is, on average, 
$\sim 0.05^m$. There is a small systematic shift of 0.03$^m$ between the two datasets, possibly due to small inaccuracies
in the photometric zero points used. Several stars that strongly deviate from the general trend are either
variable or have had their brightnesses distorted by remnants of cosmic-ray traces on the CCD frames.

\begin{table}[]
\setcaptionmargin{0mm} \onelinecaptionstrue \captionstyle{normal}
\caption{ Zero-point translation from the instrumental system to standard Lick indices (see Section~\ref{Spectroscopy}): 
$\rm I_{Lick} = a\cdot I_{measured} + b$. Columns (4) and (5) give the ranges of the indices of the standard stars used to 
carry out the calibration; the units of the indices are also given.}
\label{tab:lickSyst}
\scriptsize
\medskip
\begin{tabular}{l|c|c|c|cc}
\hline 
Index & $a, std$ & $b, std$ & Range & Unit \\ \hline
H$_{\delta_A}$ & 1.02, 0.11 & -1.03, 0.60& [-8.2, -2]& \AA \\
H$_{\delta_F}$ & 0.99, 0.06 & -0.36, 0.67& [-8.2, -2]& \AA \\
H$_{\gamma_A}$ & 0.69, 0.08 & -4.06, 0.72& [-12, 0]  & \AA \\
H$_{\gamma_F}$ & 0.49, 0.08 & -1.66, 0.23& [-4, -0.5]& \AA \\
CN1 & 1.04, 0.07 & -0.007, 0.02   & [0.1, 0.4] & mag \\
CN2 & 1.12, 0.07 & -0.03, 0.02    & [0.07, 0.4]& mag \\
Ca4227 & 1.20, 0.16 & 0.22, 0.21  & [-1.8, 3]] & \AA \\
Fe4384 & 1.05, 0.11 & 0.31, 0.61  & [2, 9.3]   & \AA \\
Ca4455 & 1.14, 0.15 & 0.43, 0.24  & [0, 2.8]   & \AA \\
Fe4531 & 1.08, 0.20 & 0.17, 0.75  & [1, 5.4]   & \AA \\
Fe4668 & 1.13, 0.06 & -0.50, 0.58 & [4, 17]    & \AA \\
H{$_\beta$}& 0.81, 0.16 & 0.22, 0.21 & [0.5, 2]   & \AA \\
Fe5015 & 1.13, 0.07 & 0.05, 0.38  & [0, 8]        & \AA \\
Mgb    & 0.80, 0.04 & 1.08, 0.13  & [0, 6]        & \AA \\
Mg$_1$ & 0.96, 0.07 & 0.03, 0.01  & [0.05, 0.25]  & mag \\
Mg$_2$ & 1.04, 0.06 & 0.03, 0.01  & [0.1, 0.4]    & mag \\
Fe5270 & 1.43, 0.19 & -0.34, 0.54 & [1, 4.5]      & \AA \\
Fe5335 & 1.09, 0.11 & 0.28, 0.28  & [0.5, 4]      & \AA \\
Fe5406 & 1.22, 0.06 & -0.18, 0.12 & [0.2, 3]      & \AA \\
Fe5782 & 1.25, 0.09 & -0.07, 0.09 & [-0.1, 1.5]   & \AA \\
Na5895 & 0.95, 0.03 & 0.23, 0.10  & [0.2, 5.2]    & \AA \\
TiO$_2$& 0.95, 0.08 & -0.01, 0.004 & [-0.05,0.1]  & mag \\
\hline
\end{tabular}
\end{table}
With some caution, since the main-sequence turn-off point is not visible in Fig.~\ref{cmd} we can conclude
from Fig. 2 that the isochrone of \cite{Bertelli} corresponding to an age of 10 billion years and a metallicity of
$\rm Z=0.001$ ($\rm [Fe/H]\sim -1.33$, according to \cite{Bertelli94}) best describes the horizontal branch of the cluster CMD.
In contrast to many other isochrones, the isochrones of \cite{Bertelli} include the horizontal-branch stage, and
enable studies of how the temperature and luminosity ranges of the horizontal-branch stars vary with their
ages, metallicities, and helium abundances. It is primarily the helium abundance, as well as the age
and metallicity, that most influence the morphology of the horizontal branch \cite{g10}. The values of the
distance modulus $\rm (m-M)_0=19.82$ ($\rm Dist.=91.9$~kpc) and color excess \mbox{$\rm E(B-V)=0.07$} we have obtained
via our fitting of the observed cluster CMD using theoretical isochrones are in reasonable agreement
with data from the literature (Table~\ref{propGCs}).

\begin{table*}[]
\setcaptionmargin{0mm} \onelinecaptionstrue \captionstyle{normal}
\caption{Lick indices ($\lambda \leq5015$\AA\ ) measured in the integrated spectrum for several stars in the central 
region of Pal~3 (Fig.~\ref{slits}).}
\label{tab:lickind1}
\begin{tabular}{lc|c|c|c|c|c|c|c|c|cc} \\ \hline                                                                                                                             
& H$_{\delta_{\rm A}}$    & H$_{\delta_{\rm F}}$    & CN$_1$                    & CN$_2$                   & Ca4227                &Ca4455                 & Fe4531                 & Fe4668               & H$_{\beta}$       & Fe5015                    \\ 
&  (\AA)               & (\AA)                & (mag)                     & (mag)                    & (\AA)                 & (\AA)                 & (\AA)                  & (\AA)                &  (\AA)              & (\AA) \\ \hline 
&1.53$^{+1.03}_{-1.04}$&0.91$^{+0.74}_{-0.75}$&-0.097$^{+0.027}_{-0.088}$ & 0.040$^{+0.010}_{-0.075}$& 0.39$^{+0.39}_{-0.40}$&0.49$^{+0.37}_{-0.37}$ & 1.71$^{+0.40}_{-0.40}$ &0.74$^{+0.49}_{-0.49}$&0.97$^{+0.22}_{-0.22}$& 1.88$^{+0.31}_{-0.31}$\\
\hline
\end{tabular}
\end{table*}
\begin{table*}[]
\setcaptionmargin{0mm} \onelinecaptionstrue \captionstyle{normal}
\caption{Lick indices   ($\lambda > 5015$\AA\ ), measured in the integrated spectrum for several stars in the central
region of Pal~3 (Fig.~\ref{slits}).}
\label{tab:lickind2}
\begin{tabular}{lc|c|c|c|c|c|c|c|c|cc} \\ \hline                                                                                                                              
 & Mg$b$                & Mg$_1$                    & Mg$_2$                   & Fe5270                & Fe5335                & Fe5406                 & Fe5782               & Na5895              & TiO$_1$  & TiO$_2$\\
&  (\AA)               & (\AA)                & (mag)                     & (mag)                    & (\AA)                 & (\AA)                 & (\AA)                  & (\AA)                &  (mag)  & (mag)\\ \hline 
&1.31$^{+0.19}_{-0.20}$& 0.009$^{+0.005}_{-0.019}$ &0.075$^{+0.005}_{-0.022}$ &0.78$^{+0.20}_{-0.20}$& 0.48$^{+0.20}_{-0.20}$ & 0.89$^{+0.15}_{-0.15}$ &0.16$^{+0.11}_{-0.11}$&1.31$^{+0.12}_{-0.12}$& 0.032$^{+0.008}_{-0.006}$& 0.004$^{+0.002}_{-0.001}$ \\
\hline
\end{tabular}
\end{table*}

\section{SPECTROSCOPY RESULTS}
\label{Spectroscopy}
\begin{figure*}[]
 \setcaptionmargin{5mm} \onelinecaptionstrue \captionstyle{normal}
\hspace{-0.9cm}
 \begin{tabular}{p{0.5\textwidth}p{0.5\textwidth}}
\includegraphics[scale=0.45,angle=-90]{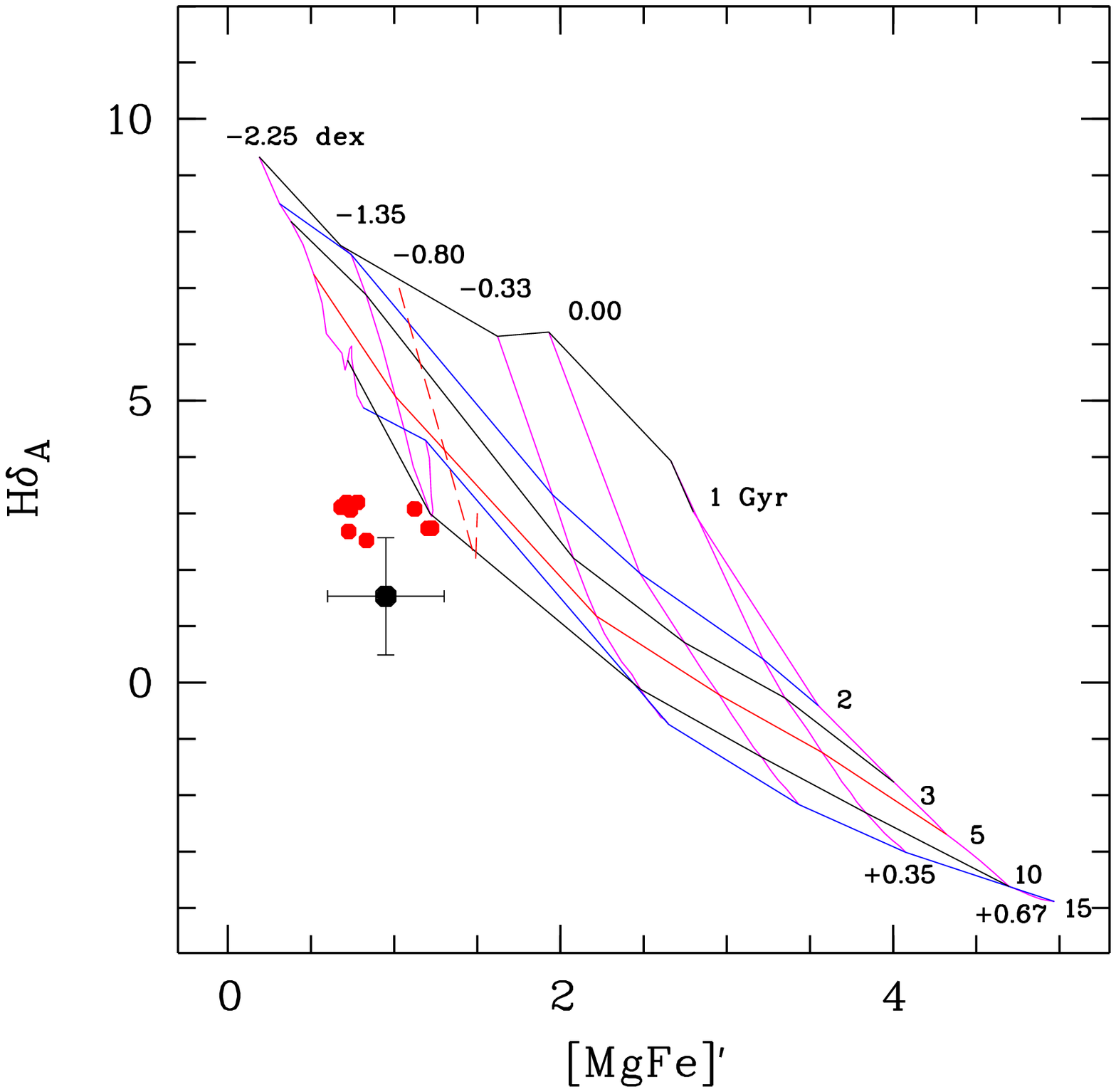} &
\includegraphics[scale=0.45,angle=-90]{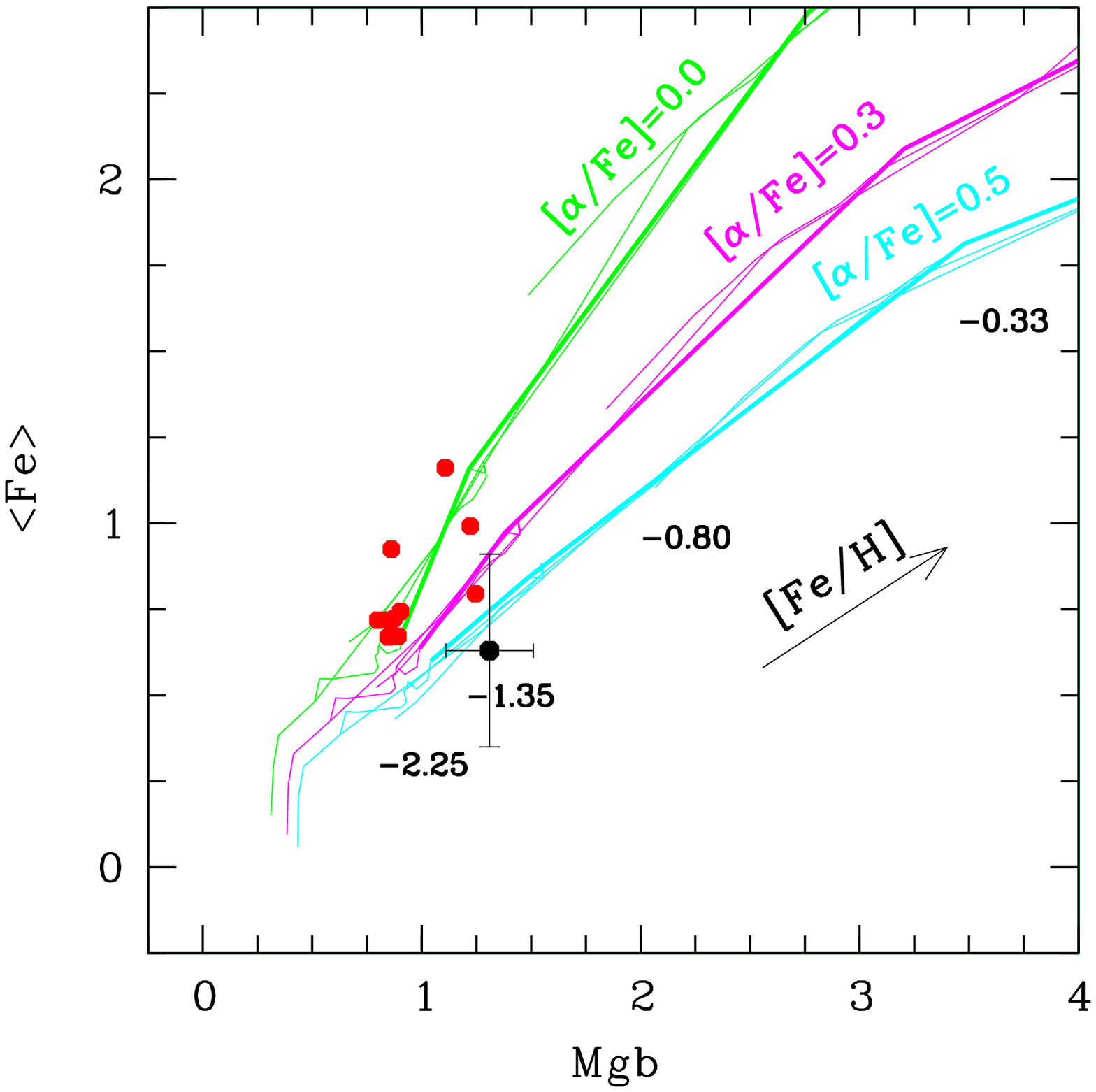} \\
\end{tabular}
 \caption{Age-metallicity (left) and metallicity-abundance of $\alpha$-elements (right) diagnostic diagrams of the Lick indices for
Pal~3 (filled circles) and Galactic globular clusters (red points) with indices from \cite{Schiavon12} and with $\rm [Fe/H]\sim -1.6$~dex (NGC~1904,
3201, 5286, 5946, 5986, 6254, 6333, 6752, and 7089) and $\rm [Fe/H]\sim -1.3$~dex (NGC~5904, 5946, 6218, 6235). The curves
show models for simple stellar populations \cite{t03}, \cite{t04}.} 
\label{fig:DiagD}
\end{figure*}
\begin{figure*}[]
 \setcaptionmargin{5mm} \onelinecaptionstrue \captionstyle{normal}
\includegraphics[scale=1.7]{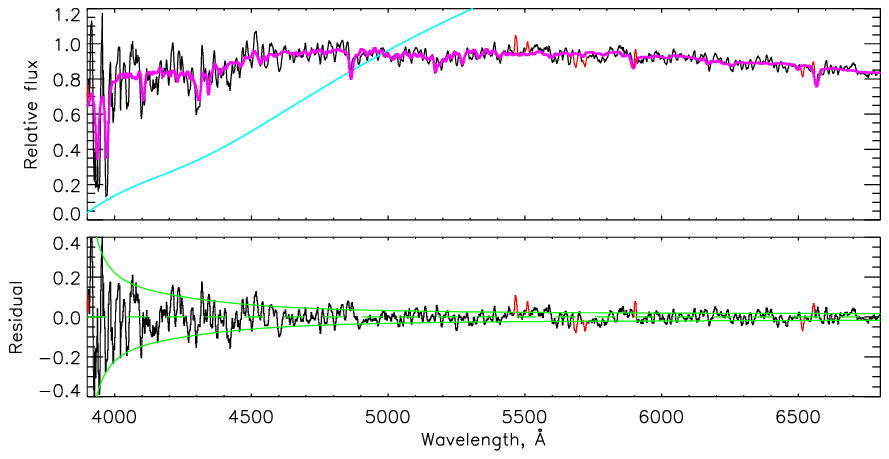} 
 \caption{Upper panel: summed stellar spectrum for Pal~3 compared with a best-fit model spectrum (bold purple curve). The
blue curve shows the results of dividing the observed spectrum by the model spectrum. Lower panel: difference between the
observed and model spectra. The green lines show the envelope corresponding to $\rm S/N=30$. The dotted line shows the zero
level. Sections of the observed spectrum that strongly deviate from the model spectrum are delineated in red.}
\label{fig:orbitPal3}
\end{figure*}
The Lick system of absorption spectral indices\footnote{\url{ http://astro.wsu.edu/worthey/html/index.table.html}}
(\cite{b_84}, \cite{Worthey94}, \cite{w94}, \cite{wo97}, \cite{t98})
 is widely used to distinguish the influence of age and metallicity on integrated spectra of repre-
sentative of old stellar populations (with ages of order a billion years or more). The Lick system is based
on indices measured in the spectra of standard stars\footnote{\url{ http://astro.wsu.edu/worthey/html/system.html, 
astro.wsu.edu/ftp/WO97/export.dat}}. Measurements of absorption line indices are reduced
to the standard system by observing objects from a list of standard stars and constructing the relationship
between the instrumental and standard indices for these stars. The coefficients of these relationships for
our observations are presented in Table~\ref{tab:lickSyst}. Tables~\ref{tab:lickind1} and \ref{tab:lickind2} 
present measurements of the Lick indices in the integrated spectra of stars in the central region of
Pal~3, reduced to the standard system. Unfortunately, because the signal-to-noise ratio per resolution
element is only $\rm S/N \sim 30$ in our spectra at the most sensitive wavelength of 5000\AA\,
the uncertainties in the measured indices are large. Our measured indices for the hydrogen lines
H$_{\beta}$, H$_{\delta}$ and H$_{\gamma}$ are
too weak compared to the corresponding values for globular clusters with similar metallicities and ages
(see, e.g., \cite{Schiavon12}, \cite{Sharina17}). One possible reason for this is a systematic overestimation of the continuum level due
to the contribution of night-sky absorption lines. We note, however, that the index ratio H$_{\delta_{\rm F}}/$H$_{\beta}=0.94$
unambiguously demonstrates that the cluster has a red horizontal branch \cite{Sch04}.

Schiavon et al. \cite{Sch04} found that Galactic clusters with blue horizontal branches ($\rm HBR\sim1$) have
higher values for this ratio than those for objects with red horizontal branches ($\rm HBR\sim-1$), 
on average, by $\sim$0.15. Further, the indices and combinations of indices in Tables~\ref{tab:lickind1} and \ref{tab:lickind2}
that are sensitive to variations in the metal abundances -- $$\langle Fe\rangle=(Fe5270+Fe5335)/2=0.63\pm0.28,$$ 
$$\rm[Mg/Fe]\arcmin=\sqrt{(Mgb\cdot(0.72 Fe5270+ 0.28 Fe5335))}$$ $=0.95\pm$0.35, Mg$b$, Mg$_1$, Mg$_2$ and others -- 
have values characteristic for Galactic clusters with $\rm [Fe/H]\sim-1.4\div-1.6$~dex.

So-called diagnostic diagrams (see, e.g., \cite{Sharina17} and references therein), two of which are shown in Fig.~\ref{fig:DiagD},
enable a comparison of the Lick indices for globular clusters with metallicities $\rm [Fe/H]\sim-1.6$~dex and $\rm [Fe/H]\sim-1.3$~dex
with the models of Thomas et al. \cite{t03}, \cite{t04}. These models for simple stellar populations
contain Lick indices computed for various ages and metallicities. The Lick indices of Pal~3 that are sensitive 
to variations in the metal content correspond to the model with $[Fe/H]\sim-1.35\pm0.2$~dex and an age
of $T=11\pm2$billion years. The results of \cite{Koch}, \cite{Hilker} and the diagram in the right panel of Fig.~\ref{fig:DiagD}
indicate a high content of $\alpha$ elements in the cluster ($\rm [\alpha/Fe]\geq 0.3$).

A comparison of the summed stellar spectrum in the central region of Pal~3 with the best-fit model
spectrum is presented in the upper panel of Fig.~\ref{fig:orbitPal3}. The best fit is provided by a simple stellar-population
model with metallicity $\rm [Fe/H]=1.35$~dex and age 10 billion years, computed by Vazdekis et al.~\cite{v10}
using the MILES spectral library \cite{sb06}. The lower panel shows the difference between the observed
and model spectra, calculated using the program {\it ULySS}\footnote{\url{ http://ulyss.univ-lyon1.fr}} (\cite{k08}, \cite{k09}). 
The use of the PEGASE.HR
grid of models \cite{l04} with the Elodie library of stellar spectra \cite{ps01} slightly increases the age and metallicity:
12$\pm2$~billion years and $\rm [Fe/H]=-1.2\pm0.25$~dex.

Thus, the results shown in this section demonstrate that fitting observed spectra with model spectra
and comparing the Lick indices with those for other Galactic clusters and with model Lick indices give
consistent results.

\section{ORBITS OF PAL 3 AND NINE MASSIVE GLOBULAR CLUSTERS}
\label{orb}
\begin{table*}[]
\setcaptionmargin{0mm} \onelinecaptionstrue \captionstyle{normal}
\caption{ Detailed characteristics for Pal~3 and nine Galactic globular clusters: (2-4) right ascension and declination at
epoch J2000.0 and corresponding uncertainties in milliarcsec (mas), (5, 6) proper motions, (7) distance from the Sun, (8)
radial velocity, (9) mass, (10) half-luminosity radii. The literature sources for data in this table are indicated 
in Sections~\ref{orb} and \ref{method_orb}.}
\label{tab:orbits9GCs}
\scriptsize
\medskip
\begin{tabular}{l|c|c|c|c|c|c|c|c|c}
\hline \hline
Object  &       RA    &       DEC   &   $\delta$RA* $\delta$DEC & $\mu_a$* & $\mu_d$       & Dist$_{\sun}$   &  $V_r$         & M            & $ r_h$\\
        & [hh mm ss]  & [gg mm ss]  &    [mas]    & [mas/yr]               &  [mas/yr]     &        [kpc]    &  [km/s]        & [M$_{\sun}$] &  [pc]   \\   \hline
Pal3    & 10 05 31.9  & +00 04 18.0 & 2200 2100   & 0.33$\pm$0.23          & 0.30$\pm$0.31 &  91.90$\pm$9.19 &   83.4$\pm$8.4 & 6.38e4       & 17.8 \\
NGC1904 & 05 24 11.09 & -24 31 29.0 & 1000 1000   & 0.20$\pm$0.43          & 1.13$\pm$0.43 &  12.89$\pm$1.42 &  205.8$\pm$0.4 & 3.57e5       & 3.00 \\
NGC5139 & 13 26 47.28 & -47 28 46.1 &  100  100   &-6.01$\pm$0.25          &-5.02$\pm$0.25 &   5.20$\pm$0.57 &  232.1$\pm$0.1 & 2.64e6       & 6.44\\
NGC5286 & 13 46 26.81 & -51 22 27.3 &  100  100   &-6.24$\pm$0.83          &-3.68$\pm$0.83 &  11.71$\pm$1.29 &   57.4$\pm$1.5 & 4.80e5       & 2.21\\
NGC6254 & 16 57 09.05 & -04 06 01.1 &  100  100   &-6.88$\pm$0.36          &-8.28$\pm$0.36 &   4.41$\pm$0.49 &   75.2$\pm$0.7 & 2.25e5       & 2.32\\
NGC6273 & 17 02 37.80 & -26 16 04.7 & 1000 1000   &-3.89$\pm$0.43          &-1.48$\pm$0.43 &   8.79$\pm$0.97 &  135.0$\pm$4.1 & 1.56e6       & 3.13\\
NGC6656 & 18 36 23.94 & -23 54 17.1 &  800  800   & 4.72$\pm$0.34          &-3.59$\pm$0.34 &   3.20$\pm$0.35 & -146.3$\pm$0.2 & 5.36e5       & 3.03\\
NGC6681 & 18 43 12.76 & -32 17 31.6 &  100  100   & 3.84$\pm$0.47          &-5.78$\pm$0.47 &   9.00$\pm$0.99 &  220.3$\pm$0.9 & 1.89e5       & 2.43\\
NGC6752 & 19 10 52.11 & -59 59 04.4 &  100  100   &-4.79$\pm$0.62          &-4.84$\pm$0.62 &   4.00$\pm$0.44 &  -26.7$\pm$0.2 & 3.64e5       & 2.72\\
NGC7089 & 21 33 27.02 & -00 49 23.7 &  100  100   & 6.77$\pm$0.42          &-4.54$\pm$0.42 &  11.51$\pm$1.27 &   -5.3$\pm$2.0 & 8.81e5       & 3.11\\
\hline
\end{tabular}
\end{table*}
\begin{figure*}[]
 \setcaptionmargin{5mm} \onelinecaptionstrue \captionstyle{normal}
 \begin{tabular}{p{0.32\textwidth}p{0.32\textwidth}p{0.32\textwidth}}
\includegraphics[scale=0.9]{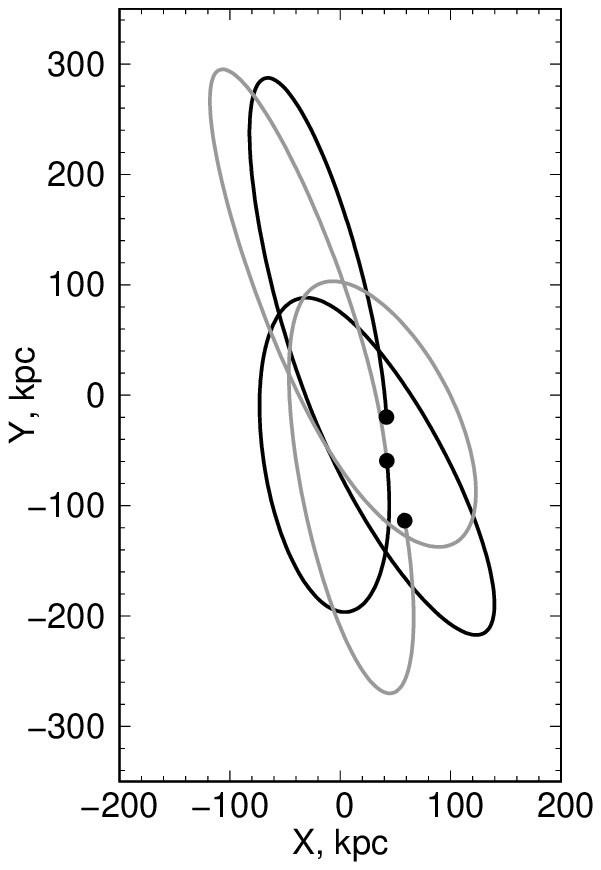} &
\includegraphics[scale=0.9]{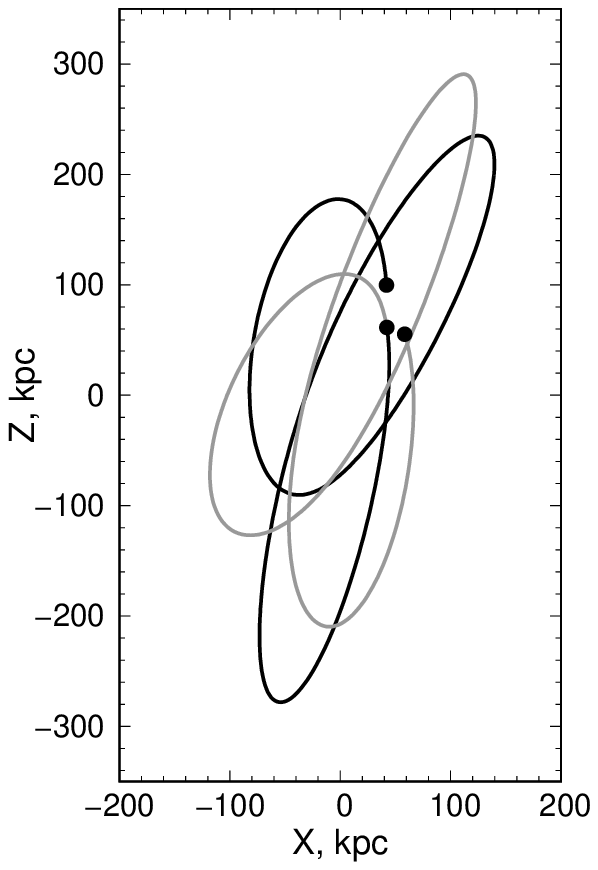} &
\includegraphics[scale=0.9]{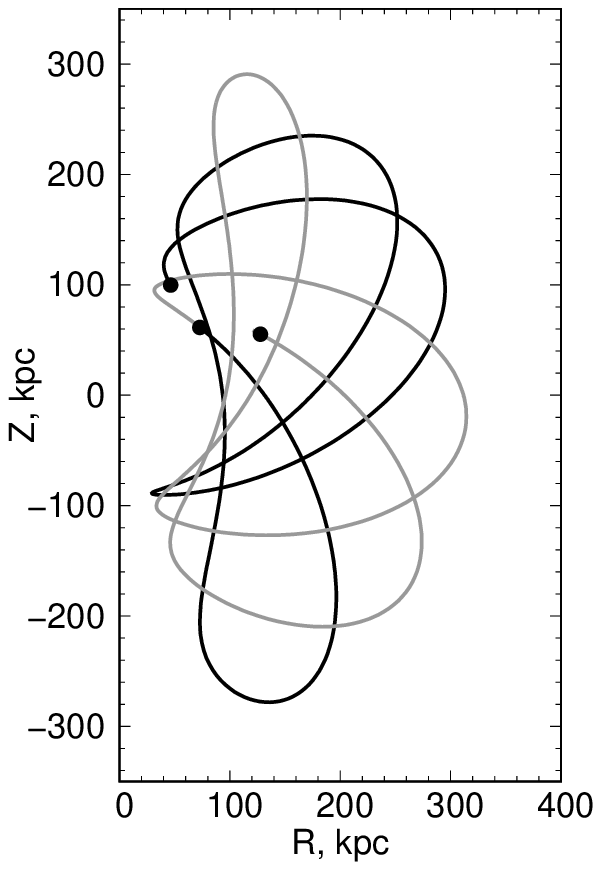} \\
\end{tabular}
 \caption{Model orbit for Pal~3 shown in projection onto the X, Y , Z, and R axes in Galactocentric coordinates. X, Y , and Z are
directed toward the center, in the direction of rotation, and toward the north Galactic pole, respectively. The current position of
the object, initial position 12 billion years in the past (black curve), and final position 12 billion years in the future (gray curve)
are marked by points.}
\label{fig:orbitPal3}
\end{figure*}
\begin{table*}[]
\setcaptionmargin{0mm} \onelinecaptionstrue \captionstyle{normal}
\caption{Parameters of the orbits of Pal~3 and nine other globular clusters with similar metallicities and ages: (2) mean
rotation period in millions of years, (3) mean orbital inclination in degrees, (4) and (5) mean radii of pericenter and
apocenter in kpc, (6) maximum distance of the cluster in the directions of the Galactic poles, $Z_{max}$, in kpc, (7) eccentricity,
(8) time required to reach the orbit pericenter and apocenter in millions of years, and (9) time required to reach the cluster's
maximum distance from the Galactic plane in millions of years. A negative mean rotation period, $T_{rot}$, denotes rotation in
the direction opposite to the direction of the Galactic rotation.}
\label{tab:orbitPal3}
\medskip
\begin{tabular}{l|c|c|c|c|c|c|c|c}
\hline \hline 
Object    & $T_{rot}$ & $Incl$     & $R_{apo}$    &   $R_{peri}$  & $Z_{max}$    & $ecc$          & $T_{ap}= T_{per}$ & $T_{Zmax}$ \\
           & (2)       & (3)        & (4)          &   (5)         & (6)          & (7)            & (8)               & (9)           \\
\hline                                                                                                                 
Pal~3  & 12056$\pm$7425& 72$\pm$9   &272$\pm$135   &   86$\pm$15   &182$\pm$101   & 0.5$\pm$0.2    & 3584$\pm$1498     & 12873$\pm$7425 \\
NGC1904 & 512$\pm$57   & 24$\pm$2   & 25$\pm$3     &  12$\pm$2     & 6.8$\pm$1    & 0.34$\pm$0.03  &  344$\pm$42       &  899$\pm$124 \\
NGC5139 &-141$\pm$3.5  & 34$\pm$2   & 6.5$\pm$0.1  & 1.7$\pm$0.2   & 1.4$\pm$0.3  & 0.59$\pm$0.03  &  93$\pm$2         &  164$\pm$14   \\
NGC5286 & 333.5$\pm$97 & 32$\pm$6.5 & 15$\pm$6     & 8$\pm$1       & 5.3$\pm$2.2  & 0.29$\pm$0.11  & 216$\pm$64        &  552$\pm$198   \\
NGC6254 & 123$\pm$6.6  & 57$\pm$8   & 5.4$\pm$0.3  & 1.5$\pm$0.16  & 2.5$\pm$0.4  & 0.57$\pm$0.03  &  83$\pm$4         &  192$\pm$9  \\
NGC6273 & -37.8$\pm$38 & 71$\pm$12  & 2.3$\pm$0.8  & 0.9$\pm$0.2   & 1.4$\pm$0.2  & 0.42$\pm$0.21  &  35$\pm$12        &  86$\pm$13  \\
NGC6656 & 187$\pm$6.7  & 17$\pm$2   & 8.9$\pm$0.3  & 2.9$\pm$0.2   & 0.8$\pm$0.1  & 0.51$\pm$0.02  &  120$\pm$4        &  164$\pm$7  \\
NGC6681 & 116$\pm$209  & 82$\pm$4   & 11$\pm$3     & 1.3$\pm$0.4   & 5$\pm$1      & 0.76$\pm$0.11  &  135$\pm$28       &  359$\pm$60 \\
NGC6752 & 153$\pm$7    & 35$\pm$4   & 5.8$\pm$0.2  & 3.4$\pm$0.4   & 2.1$\pm$0.2  & 0.27$\pm$0.05  &  99$\pm$4         &  207$\pm$8 \\
NGC7089 &-815$\pm$285  & 69$\pm$2   & 45$\pm$17    & 6.7$\pm$1.2   & 25$\pm$8     & 0.73$\pm$0.04  &  514$\pm$182      &  1596$\pm$568  \\
\hline
\end{tabular}
\end{table*}

\begin{figure*}[]
 \setcaptionmargin{5mm} \onelinecaptionstrue \captionstyle{normal}
 \begin{tabular}{p{0.34\textwidth}p{0.34\textwidth}p{0.34\textwidth}}
\includegraphics[scale=0.6]{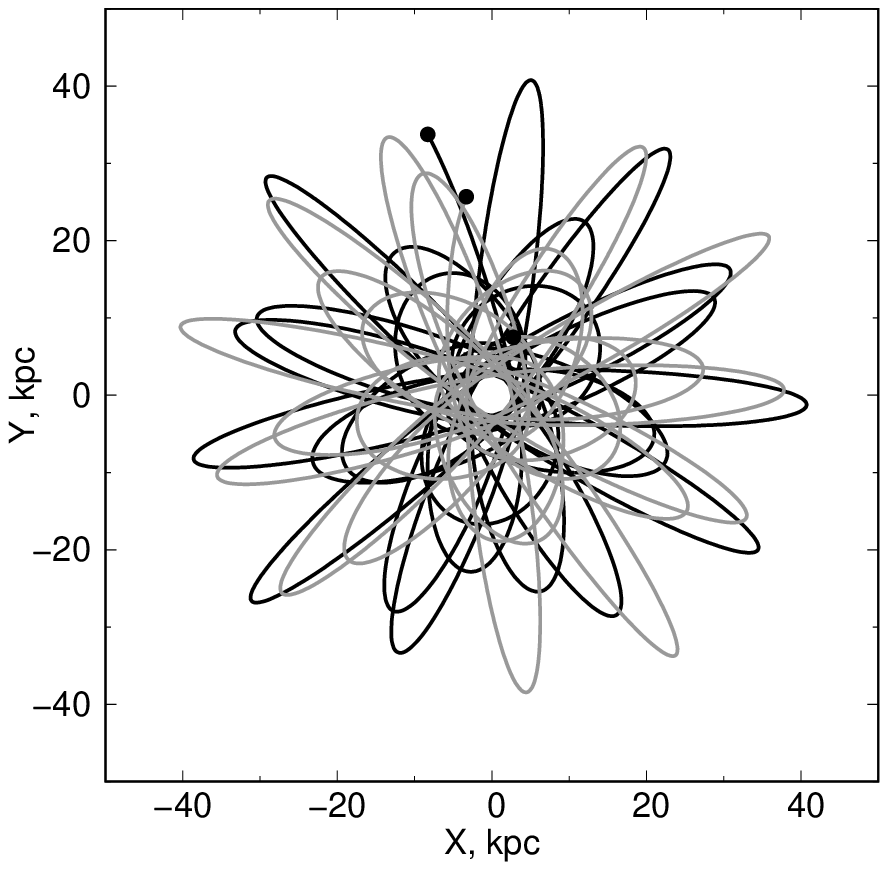} &
\includegraphics[scale=0.6]{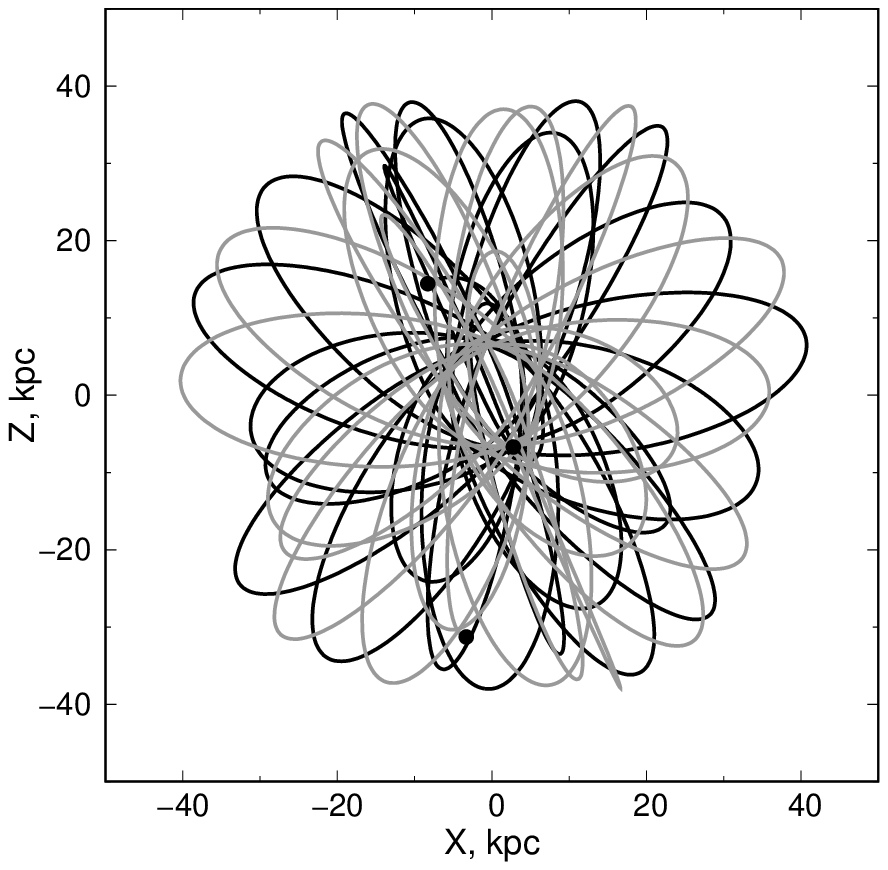} &
\includegraphics[scale=0.6]{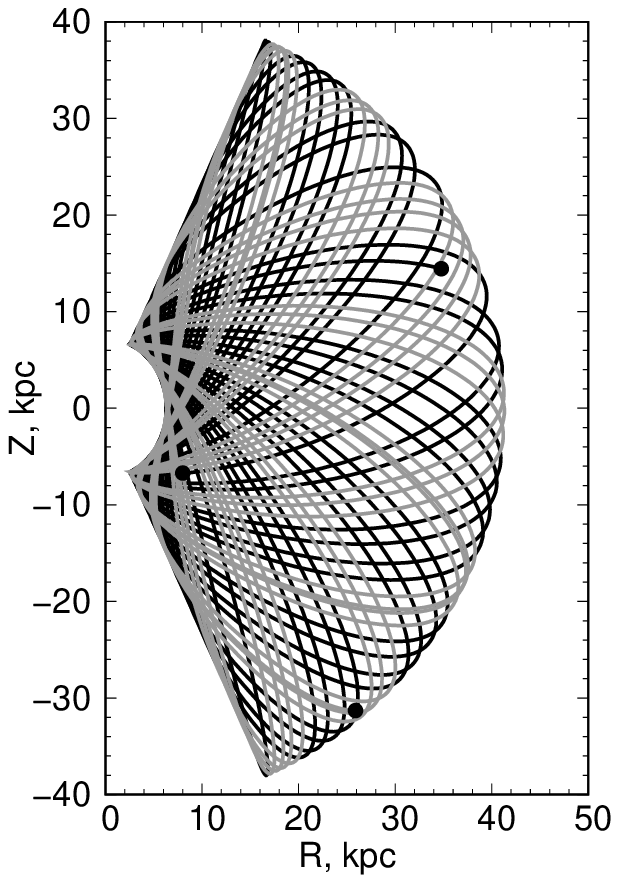} \\
\end{tabular}
 \caption{Model orbit for NGC~7089 shown in projection onto the X, Y , Z, and R axes in Galactocentric coordinates. X, Y ,
and Z are directed toward the center, in the direction Galactic rotation, and toward the north Galactic pole, respectively. The
current position of the object, initial position 12 billion years in the past (black curve), and final position 12 billion years in the
future (gray curve) are marked by points.}
\label{fig:orbitN7089}
\end{figure*}

In this section, we consider the orbit of Pal~3 and compare it with the orbits of several massive Galactic
globular clusters with similar metallicities and ages, $\rm [Fe/H]\sim-1.6$~dex, $\rm T\sim11.2$~billion years (\cite{Sharina17} and
references therein): NGC~1904, 5139, 5286, 6254, 6273, 6656, 6681, 6752, and 7089. The main charac-
teristics of the clusters related to the computation of their orbits are given in Table~\ref{tab:orbits9GCs}. The cluster masses
from \cite{go97} and half-luminosity radii from \cite{mvdb05} are also given.

\subsection{Method for Orbit Computation}
\label{method_orb}
Data on distances, radial velocities, and the proper-motion components were taken from the
catalogs of Harris \cite{Harris}\footnote{online version: \\
\url{http://physwww.mcmaster.ca/~harris/mwgc.dat}} and Kharchenko et al. \cite{Kharchenko}.
The proper motions for Pal~3 were taken from \cite{mc93}.
The positions of the clusters (apart from Pal~3, NGC~1904, and NGC~6273) were taken from \cite{Goldsbury}.
The positions of Pal~3, NGC~1904, and NGC~6273 were taken from \cite{Cotton} (Pal3 = UGC 05439),
\cite{Lanzoni}, and \cite{Picard}, respectively.

We considered an axially symmetrical potential for the Galaxy consisting of potentials for the disk,
bulge, and halo. The potential of the bulge $\Phi_b(r)$ was described using a spherically symmetrical Plummer
distribution \cite{Plummer1911},
 \begin{equation}
  \Phi_b(r)=-\frac{M_b}{(r^2+b_b^2)^{1/2}},
  \label{bulge}
 \end{equation}
 and the potential of the disk $\Phi_d(R,Z)$ was taken to have the form proposed by Miyamoto and Nagai \cite{Miyamoto}:
 \begin{equation}
 \Phi_d(R,Z)=-\frac{M_d}{\Biggl[R^2+\Bigl(a_d+\sqrt{Z^2+b_d^2}\Bigr)^2\Biggr]^{1/2}},
 \label{disk}
\end{equation}
where $M_b$ and $M_d$ are the mass of the bulge and disk, and $b_b$, $a_d$, and $b_d$ are scaling parameters of the
components in kpc. We used a logarithmic potential to describe the dark-matter halo \cite{Rubin80}:
 \begin{equation}
 \Phi_h(r) = \frac{v_0^2}{2} \ln(r^2+R_h^2)
 \label{halo}
\end{equation}
with the asymptotic rotational velocity $v_0 = 220$~km/s and radius $R_h$ , determined from the condition that
the circular velocity in the plane of the disk attain the value $v_0$ at $R = 8.5$~kpc when the full potential is
taken into account, i.e., including both the bulge~(\ref{bulge}) and the disk~ (\ref{disk}).

The parameters of the bulge and disk potentials (distance from the Sun to the Galactic center, lo-
cal circular rotational velocity of the Galaxy, scaling parameters of the disk, bulge, and halo) were taken
from \cite{Allen}.

The translation from the coordinates in the International Celestial Reference Frame to Galactocentric
coordinates was done using a rotation matrix based on fitting of a catalog of objects observed by 
Hipparcos \cite{ESA}. The position and velocity of the Sun that we used were
$X_\odot = 8.33$~kpc \cite{Gillessen}, $Z_\odot = 27$~pc \cite{Chen2001},
$(U,V,W)_\odot = (11.1, 12.24, 7.25)$~km/s \cite{Schonrich}, $V_{LSR} = 218$~km/s \cite{Bovy2012}.

The cluster orbits were characterized by the mean parameter values and their deviations for 1000 
versions of the orbits computed 12 billion years into the past and 1000 versions computed 12 billion years into
the future. The integration was carried out using a 5th-order Runge-Kutta-Feldberg method with a
variable step \cite{Fehlberg}. The input parameters were randomly varied 1000 times in accordance with 
the uncertainties in the observational parameters (the set of initial conditions for each cluster was generated based
on a normal distribution bounded by  $2\sigma$; we did not include systematic errors associated with uncertainty
in the position and velocity of the Sun relative to the Galactic center). It is important that the acceleration
of the cluster in the Galactic potential depends on the cluster mass only in the case of dynamical friction. 
This effect is usually important only for massive globular clusters whose trajectories pass close to the
Galactic center. We did not include the effect of dynamical friction in our study. The results of con-
structing model orbits for Pal~3 and nine other Galactic clusters are presented in Table~\ref{tab:orbitPal3}. The models of
the orbits are shown graphically in Fig.~\ref{fig:orbitPal3} for Pal~3, in Fig.~\ref{fig:orbitN7089} for NGC~7089, 
and at the ftp site \url{ftp://ftp.sao.ru/pub/sme/V220_plot/} for the other clusters.

\subsection{Results of Orbit Computations}
The tables and figures illustrating the computed orbit models show that the orbit of Pal~3 clearly does
not resemble any of the orbits of the other nine massive Galactic clusters with similar ages and metallicities, 
in terms of either its shape or its characteristics. Pal~3 is moving around the Galactic center along a
complex trajectory. The period for the cluster to reach its maximum distance from the Galactic center is
comparable to the age of the Universe. The cluster emerged from a position close to this most distant
point roughly 12 billion years ago, and the object is currently located near the pericenter of its orbit.
We obtained a lower value for the maximum distance of Pal~3 than was found in \cite{Palma}. According to our
computations, this distance is about 340~kpc. The cluster has approached the Galactic center to a min-
imum distance of 70~kpc. Therefore, only relaxation as a result of stellar encounters exerts an important
influence on the dynamical and structural evolution of this cluster \cite{go97}. The effects of dynamical friction and
collisions with the disk and bulge are not important due to the cluster's appreciable distance from the
Galactic plane.

The orbits of the other nine clusters are close to the Galactic plane (in the Z direction, Table~\ref{tab:orbitPal3}). 
In contrast to Pal~3, these nine clusters are some of the most massive in the Galaxy. They have high central
densities and unusual properties for their stellar populations, such as extremely blue horizontal branches,
multiple evolutionary sequences, and anticorrelations between elemental abundances (see \cite{Sharina17} 
and references therein). Some of these objects correspond to stellar flows. The clusters have probably passed
through dense gaseous layers many times, becoming closer and closer to the Galactic center and the
Galactic plane due to the effects of dynamical friction and collisions with the disk and bulge \cite{go97}. The high
masses and high central densities of these clusters have hindered their total disruption during the course
of their evolution.

The parameters of the orbit models for the massive clusters from the literature overall resemble our own
values (\cite{Dinescu}, \cite{Balbinot} and \cite{Pouliasis}). 
Dinescu et al. \cite{Dinescu} and Balbinot and Gieles \cite{Balbinot} also computed elements of the orbit
of Pal~3. The differences are larger for more distant clusters, for which the uncertainties in their proper
motions and other parameters are higher. The mean rotation periods of four of the clusters have large
uncertainties: Pal~3, NGC~5286, NGC~6681, and NGC~7089. The case of Pal~3 is the most complex. Its
proper motion is known only with large uncertainties: $\mu_{a}*=0.33\pm0.23$, $\mu_d=0.30\pm0.31$ \cite{mc93}. 
Differences in the methods used to construct the models and large uncertainties in the observed parameters lead
to differing orbital parameters. Dinescu et al. \cite{Dinescu} used the proper motions of \cite{mc93} and obtained the
following elements of the orbit: mean rotation period in units of $10^6$~years $P\sim9740$, radius of apocenter
$R_{apo}\sim420$~kpc, radius of pericenter $R_{peri}\sim83$~kpc, orbital eccentricity $ecc\sim0.7$, and maximum distance
of the cluster in the direction of the Galactic poles $Z_{max}\sim308$~kpc. Balbinot and Gieles \cite{Balbinot} also used
the proper motions of Pal~3 from \cite{mc93}, as did we. However, the radii of apocenter and pericenter and
the orbital eccentricities they obtained were $R_{apo}\sim1985$~kpc,  $R_{peri}\sim94$~kpc, $ecc\sim0.9$. Our own derived
orbital parameters are closer to the values of \cite{Dinescu}.

\section{DISCUSSION AND CONCLUSION}
\label{discussion}

We have attempted to improve our knowledge of the origin of Pal~3 -- one of the most distant globular clusters --
by analyzing various observational data available to us: direct VLT frames, integrated spectra
of stars in the central region of the cluster, and catalog information about the object's kinematics.

We carried out stellar photometry using previously unused archival VLT data, and estimated the distance
to Pal~3, $Dist.=91.9$kpc, and its color excess, \mbox{$E(B-V)=0.07$}. We have found that the isochrone of \cite{Bertelli}
corresponding to an age of 10 billion years, metallicity $Z=0.001$ ($\rm [Fe/H]\sim -1.33$, according to the 
empirical calibration of \cite{Bertelli94}), and helium abundance $Y=0.26$ fits the distribution of stars on the horizontal
branch of the cluster better than other isochrones. Deeper direct frames are needed to carry out photometry 
stars at the main-sequence turn-off point, and to refine the age, metallicity, and helium abundance
of the cluster. Deep spectral observations of stars in Pal~3 could answer the question of whether two or
three stars on the continuation of the red horizontal branch toward the blue are cluster members. This
will make it possible to refine the cluster's horizontal-branch index HBR and evolutionary parameters. The
absence of stars near the tip of the red-giant branch of Pal~3 and the closeness of the luminosities of stars at
the main-sequence turn-off point to the photometric limit has hindered precise estimation of the cluster's
age and metallicity based on its CMD.

We have carried out spectroscopic studies of stars in the central region of the cluster based on archival
data from the CARELEC spectrograph of the 1.93-m telescope of the OHP; the total exposure time for stars
in the central region of the cluster was 3.75 hours under good observing conditions. Our fitting of
a model to the observed spectrum at wavelengths [3900:6800]~\AA\ and comparison with model measurements 
of the Lick indices indicates that Pal~3 has an enhanced abundance of $\alpha$-process elements,
$\rm [\alpha/Fe]\geq0.3$, an age of $T=11\pm2$ billion years, and metallicity $\rm [Fe/H]=-1.4\pm0.2$~dex. The values of
$\rm [\alpha/Fe]$ and $T$ derived through our spectroscopic and photometric analyses are in good agreement with
data from the literature. Our value of $\rm [Fe/H]$ is about 0.2 dex higher than previously determined values in
the literature.

Our constructed model orbit for Pal~3 indicates that this cluster is located near its closest distance
from the Galactic center. In its path from the most distant point on its orbit, $R_{gc}\sim$340~kpc, the cluster 
has approached the Galactic center to a minimum distance of $\sim$70~kpc only once. Our computed
model orbit suggests a probable extragalactic origin for Pal~3.

It is difficult to determine which of the nearby existing galaxies originally hosted the cluster based
on the available observational data. The Phoenix dwarf galaxy has a transitional morphological type
between a gas-rich irregular galaxy and a decaying spheroidal galaxy (dIrr/dSph, $Dist.=0.43$~Mpc \cite{p11}),
and an orbit that is coplanar with the orbit of Pal~3, according to Palma et al.~\cite{Palma}. The stellar 
population of Phoenix is characterized by a wide range of metallicities and an appreciable metallicity gradient
(\cite{k16} and references therein). For most evolved stars in Phoenix, [Fe/H] lies in the range from -1~dex
to -2~dex. The mean metallicity of this galaxy is $\rm \langle [Fe/H] \rangle= -1.49 \pm 0.04$~dex. 
No observational data about the proper motion of Phoenix are available.

Palma et al. \cite{Palma} undertook a statistical study of the distribution of orbital moments for galaxies in the 
Local Group and globular clusters. They distinguished a number of dynamical subgroups of objects. Pal~3
is a member of the subgroup of dwarf galaxies: the Small Magellanic Cloud and Ursa Minor. In addition
to these two galaxies, this subgroup also contains the cluster NGC~5024.

Peebles et al. \cite{p11} constructed a dynamical numerical model for the orbit of the Phoenix dwarf galaxy.
The input parameters for the model were the position and radial-velocity components for galaxies in the
Local Group at distances of less than 1.5 Mpc from the Galaxy. Simulations enabled estimation of the
masses and proper motions of the galaxies. It was found that Phoenix was the first to approach this close
to the Milky Way, and its motion, like those of most other dwarf galaxies, originated in the outskirts of the
Local Group. Thus, more accurate proper-motion data should help elucidate whether or not the cluster
Pal~3 was ejected from this galaxy, and, if it was, what was the origin of this event.

Future observations using large ground and space telescopes will help clarify the origin of Pal~3, and also
of five other globular clusters with similar ages and metallicities located far from the Galactic center.

\begin{acknowledgments}
This work was supported by the Russian Foundation for Basic Research (grant nos. 17-52-45053,
18-02-00167) and the Ministry of Education and Science of Russia (project no. 3.858.2017/4.6). This
study was based on observations collected at the European Southern Observatory through the program
ESO 077.D-0775, and used services of the ESO Scientific Archive Foundation. We thank E. Davoust and
the Observatoire de Haute Provence for the spectral observations used in our study. We also thank the
anonymous referee for comments that have enabled us to improve this paper.
\end{acknowledgments}

{\it \hspace{3cm}Translated by D. Gabuzda}
\end{document}